%% file: main.tex
\documentclass[11pt]{article}
\pdfoutput=1

\usepackage[T1]{fontenc}              % T1 font encoding as default
\usepackage[latin9]{inputenc}         % ISO-8859-15 (Latin 9) text

\usepackage[a4paper]{geometry}        % DIN A4 paper

\usepackage{url}

\usepackage{booktabs}

\usepackage{graphicx} % WE USE PDFLATEX
\usepackage{amssymb,amsfonts,amsmath}

\usepackage{natbib}
\usepackage{mathpazo}
\usepackage{microtype}

\input{abbreviations}

\begin{document}

\date{4 February 2009; last revision 16 July 2009}
\title{Gene ranking and biomarker discovery under correlation}

\author{Verena Zuber \thanks{Institute for Medical Informatics,
      Statistics and Epidemiology (IMISE),
      University of Leipzig,
      H\"artelstr. 16--18,
      D-04107 Leipzig, Germany}
{} and
 Korbinian Strimmer \footnotemark[1]
}
\maketitle

\input{abstract}

\newpage

\input{body}

\bibliographystyle{apalike}
\bibliography{preamble,econ,genome,stats,array,sysbio,misc,molevol,med,entropy}

\end{document}

%% file: abbreviations.tex
\newcommand\bdelta{\boldsymbol \delta}
\newcommand\bI{\boldsymbol I}
\newcommand\bM{\boldsymbol M}
\newcommand\bmu{\boldsymbol \mu}
\newcommand\bomega{\boldsymbol \omega}
\newcommand\bP{\boldsymbol P}
\newcommand\bR{\boldsymbol R}
\newcommand\bSigma{\boldsymbol \Sigma}
\newcommand\btau{\boldsymbol \tau}
\newcommand\bt{\boldsymbol t}
\newcommand\bU{\boldsymbol U}
\newcommand\bV{\boldsymbol V}

\newcommand\bx{\boldsymbol x}
\newcommand\bZ{\boldsymbol Z}

\newcommand\prob{\text{Pr}}

\newcommand{\figcite}[1]{Fig.~\textbf{\ref{#1}}}
\newcommand{\eqcite}[1]{Eq.~\textbf{\ref{#1}}}
\newcommand{\tabcite}[1]{Tab.~\textbf{\ref{#1}}}
\newcommand{\eqscite}[2]{Eqs.~\textbf{\ref{#1}} and \textbf{\ref{#2}}}

%% file: abstract.tex
\begin{abstract}
\noindent\textbf{Motivation:}
Biomarker discovery and gene ranking is a standard task in genomic high
throughput analysis.  Typically, the ordering of markers is based on
a stabilized variant of the $t$-score, such as the moderated $t$ or the
SAM statistic.  However, these 
procedures ignore gene-gene correlations, which may have a profound
impact on the gene orderings and on the power of the subsequent tests.

\noindent\textbf{Results:}
We propose a simple procedure that adjusts 
gene-wise $t$-statistics to take account of
correlations among genes.  The resulting correlation-adjusted
$t$-scores (``cat'' scores) are  derived from a
predictive perspective, i.e. as a score for 
variable selection to discriminate group membership in two-class
linear discriminant analysis.  In the absence of correlation the cat 
score reduces to the standard $t$-score.
Moreover, using the cat score it is straightforward 
to evaluate groups of features (i.e. gene sets).
For computation of the
cat score from small sample data we propose a shrinkage procedure.
In a  comparative study comprising
 six different synthetic and empirical correlation structures 
 we show that the cat score improves estimation 
of gene orderings and leads to higher power for fixed true 
discovery rate, and vice versa.  Finally, we also illustrate the cat 
score by analyzing metabolomic data.

\noindent\textbf{Availability:}
The shrinkage cat score is implemented in the R package ``st'' 
available from URL \url{http://cran.r-project.org/web/packages/st/}.

\noindent\textbf{Contact:} \url{strimmer@uni-leipzig.de}
\end{abstract}

%% file: body.tex
\section{Introduction}

The discovery of genomic biomarkers is often based on 
case-control studies.  For instance, consider a typical microarray experiment  
comparing healthy to cancer tissue. A shortlist of genes relevant for 
discriminating the phenotype of interest is compiled by 
ranking genes according to their respective $t$-scores.
Because of the high-dimensionality of the genomic data
special stabilizing procedures such as ``SAM'',
 ``moderated $t$'' or ``shrinkage $t$'' are warranted and most effective
-- see \citet{OS07a} for a recent comparative study.

However, microarrays  are only one particularly prominent example
of a series of modern technologies  emerging for high-throughput 
biomarker discovery.  In addition to gene expression it is now common
practice in biomedical laboratories to measure metabolite concentrations 
and protein abundances.
A distinguishing feature of proteomic and metabolic data
is the presence of \emph{correlation among markers}, due to 
chemical  similarities (metabolites) and spatial dependencies 
(spectral data).  These correlations may impact statistical conclusions.

There are three main strategies for dealing with the issue
of correlation among biomarkers.  
One approach is to initially ignore the correlation structure and 
to compute conventional $t$-scores. Subsequently,
the effects of correlation are accommodated in the last stage of the analysis
 when statistical significance is assigned  \citep{Efr07a,SLW08}. 
An alternative approach is to 
model the correlation structure explicitly in the data
generating process, and base all inferences on this more complex
model.  For example, in case of proteomics data a spatial
autoregressive model can account for dependencies 
between neighboring peaks \citep{Hand08}.
A third strategy, occupying
middle ground between the two described approaches, is to combine $t$-scores
and the estimated correlations to a new gene-wise test statistic.  This approach
is followed in  \citet{TW06} and \citet{Lai08} and it is also the route
we pursue here.

Specifically, we propose  ``correlation-adjusted $t$''-scores, 
or short ``cat'' scores.  These scores  are derived from
a predictive perspective by exploiting a close link
between gene ranking and and two-class linear discriminant analysis (LDA).
It is well known \citep{FF08} that the
$t$-score is the natural feature selection criterion in diagonal
discriminant analysis, i.e.\ when there is no correlation.
As we argue here in the general LDA case assuming arbitrary 
correlation structure this role is taken over by the cat score.

For practical application of the cat score as a ranking 
criterion for biomarkers we develop a corresponding shrinkage procedure, 
which can be employed in high-dimensional settings with a comparatively 
small number of samples.  This statistic reduces to the 
shrinkage $t$ approach  \citep{OS07a}
if there is no correlation.  We also provide a recipe for constructing
cat scores from other regularized $t$-statistics.
Furthermore, we show that the cat score  enables in a straightforward fashion
 the ranking of \emph{sets} of features and thus facilitates the analysis 
of gene set enrichment \citep{AS09}.

The rest of the paper is organized as follows.  Next, we present the 
methodological background, the definition of the cat score, and a
corresponding small sample shrinkage procedure.
Subsequently, we report results from a comparative study where
we investigate the performance of the shrinkage cat score
relative to other gene ranking procedures including the approaches by \citet{TW06} and 
\citet{Lai08}.  In our study we assume a variety of both synthetic as well 
as empirical correlation scenarios from gene expression data.  
Finally, we illustrate the cat score approach by analyzing a metabolomic
data set and conclude with recommendations.

\section{Methods}

%In the following we first briefly review linear discriminant analysis and then proceed to 
%motivate the cat score both for gene ranking and feature selection.
%Subsequently, we develop a shrinkage procedure for computing the cat score
%from small sample data. Finally, we explain the use of the cat score for 
%ranking both individual genes as well as of groups of features.

\subsection{Linear discriminant analysis}

Linear discriminant analysis (LDA) is a simple yet
very effective classification algorithm \citep{Hand06}.
If there are $K$ distinct class labels, then LDA assumes
that each class can be represented by a multivariate
normal density 
\begin{equation*}
\begin{split}
f( \bx | k )  &= (2 \pi)^{-p/2} | \bSigma|^{-1/2} \; \times \\
 & \mathrel{\phantom{=}} \exp\{  -\frac{1}{2} (\bx-\bmu_k)^T \bSigma^{-1} (\bx-\bmu_k) \} 
\end{split}
\end{equation*}
with mean $\bmu_k$ and a common covariance matrix $\bSigma$, which 
can be decomposed into $\bSigma = \bV^{1/2} \bP \bV^{1/2}$  with 
 correlations $\bP=(\rho_{ij})$
and variances $\bV = \text{diag}\{\sigma^2_1, \ldots, \sigma^2_p\}$.
The  observed  $p$-dimensional data $\bx$ (e.g., the expression
levels of all genes in a sample) are thus modeled by the mixture
$$
f(\bx) = \sum_{j=1}^K \pi_j f(\bx | j) , 
$$
where the $\pi_j$ are the a priori mixing weights. Applying Bayes' theorem gives the probability of group $k$ given $\bx$, 
\begin{equation*}
\prob(k| \bx) = \frac{\pi_k f(\bx | k)}{f(\bx)} ,
\end{equation*}
which in turns allows to define the discriminant score 
$d_k(\bx) = \log\{\prob(k| \bx)\}$. 
Dropping terms  constant across groups this results
for LDA in
\begin{equation*}
\begin{split}
d_k^\text{LDA}(\bx) &=  \bmu_k^T \bSigma^{-1} \bx -\frac{1}{2} 
                         \bmu_k^T \bSigma^{-1} \bmu_k  + \log(\pi_k)  \\
	            &=  \bmu_k^T (\bV^{1/2} \bP \bV^{1/2})^{-1} \bx \\
                    &\phantom{=} -\frac{1}{2} \bmu_k^T (\bV^{1/2} \bP \bV^{1/2})^{-1} \bmu_k  + \log(\pi_k) \, .
\end{split}
\end{equation*}
Due to the common covariance $d_k^\text{LDA}(\bx)$ is linear in $\bx$,
hence the name of the procedure.
In order to assign a class label to a data sample $\bx$
the discriminant function for all classes is computed, and
the class is selected that maximizes $d_k(\bx)$.  
The discriminant 
function itself is learned from a separate training data set
(i.e.\ independently from the test samples).

An important special case of LDA is diagonal discriminant analysis (DDA),
to which LDA reduces if there is no correlation ($\bP=\bI$) among features.
Then  the  discriminant function simplifies to
$$
d_k^\text{DDA}(\bx) = 
 \bmu_k^T \bV^{-1} \bx -\frac{1}{2} 
 \bmu_k^T \bV^{-1} \bmu_k  + \log(\pi_k).
$$
In the machine learning literature prediction using the
function $d_k^\text{DDA}(\bx)$ 
is known as ``naive Bayes'' classification \citep{BL04}. 

\subsection{Feature selection in two-class LDA}

Gene ranking and feature selection for class prediction are closely connected.
We exploit this here to define a score for ranking genes (features) in the presence of 
correlation.
In what follows, we consider  LDA for precisely two classes, i.e.\
the typical setup in case-control studies.  

For $K=2$ the difference
$\Delta^\text{LDA}(\bx) = d_1^\text{LDA}(\bx) - d_2^\text{LDA}(\bx)$
 between the discriminant
scores of the two classes provides a simple prediction rule:
if $\Delta^\text{LDA} \geq  0$ then a test sample is assigned to class 1, 
otherwise  class 2 is chosen.
$\Delta^\text{LDA}(\bx)$ can be written after some algebra
\begin{equation}
\Delta^\text{LDA}(\bx) = 
 \bomega^T \bdelta(\bx)
+ \log(\frac{\pi_1}{\pi_2})
\label{eq:twoclasslda}
\end{equation}
with weight vector
\begin{equation}
\bomega =  \bP^{-1/2} \bV^{-1/2} ( \bmu_1 -  \bmu_2 )
\label{eq:featureweights}
\end{equation}
and vector-valued distance function
\begin{equation}
\bdelta(\bx) = \bP^{-1/2} \bV^{-1/2}  ( \bx - \frac{ \bmu_1+\bmu_2}{2}).
\label{eq:distfunc}
\end{equation}
The benefit of expressing two-class LDA in this fashion is that 
it clarifies the underlying mechanism. In particular,
the difference score $\Delta^\text{LDA}(\bx)$ is governed solely by three factors:
\begin{itemize} 
\item the log-ratio of the mixing proportions $\pi_1$ and $\pi_2$,
\item $\bdelta(\bx)$, the standardized and decorrelated distance  of the
      test sample $\bx$ to the average centroid, and
\item the variable-specific feature weights $\bomega$.
\end{itemize}
Note in particular that the weight vector $\bomega$ is not a function of the test data 
$\bx$ and that it carries no  units of measurements. 
Its components $\omega_i$ directly control how much each particular
 gene $i$ contributes to the overall score $\Delta^\text{LDA}$.
Thus,  $\bomega$ is a natural univariate indicator for feature 
selection in  two-class linear discriminant analysis.
Moreover, note that the function $\bdelta(\bx)$ is a Mahalanobis transform,
i.e. the predictors $\bx$ are centered, standardized and sphered before feature selection.

The interpretation of $\bomega$ as general feature weights is supported by considering the
special case of DDA. In the absence of correlation the weights
$\bomega$ directly reduce to $\bV^{-1/2} ( \bmu_1 -  \bmu_2 )$, which 
is (apart from a constant) the usual vector of two-sample $t$-scores.
It is well known that  in the DDA setting the $t$-score 
is the natural and optimal ranking criterion
for discovering genes that best differentiate the two classes \citep{FF08}.
Note that if we would (hypothetically) decompose the product $\bomega^T \bdelta(\bx)$
from \eqcite{eq:twoclasslda} in a different
fashion, e.g., such that the factor $\bP^{-1/2} \bV^{-1/2}$
was moved from
\eqcite{eq:distfunc} to \eqcite{eq:featureweights}, then in the limit
of vanishing correlation the ranking criterion would 
\emph{not} be a $t$-score but rather  $\bV^{-1} ( \bmu_1 -  \bmu_2 )$. 
Similarly, if we would move only the 
factor $\bP^{-1/2}$ from \eqcite{eq:distfunc} to \eqcite{eq:featureweights},
then a number of other inconsistencies arise, in particular the
connection of $\bomega$ with Hotelling's $T^2$ statistic (see further below) 
is lost.  Therefore, the decomposition as given by 
 \eqcite{eq:featureweights} and \eqcite{eq:distfunc} is the most natural.

\subsection{Definition of the correlation-adjusted $t$-score (cat score)}

\begin{figure*}[t]
\begin{center}
\centerline{\includegraphics[width=15cm]{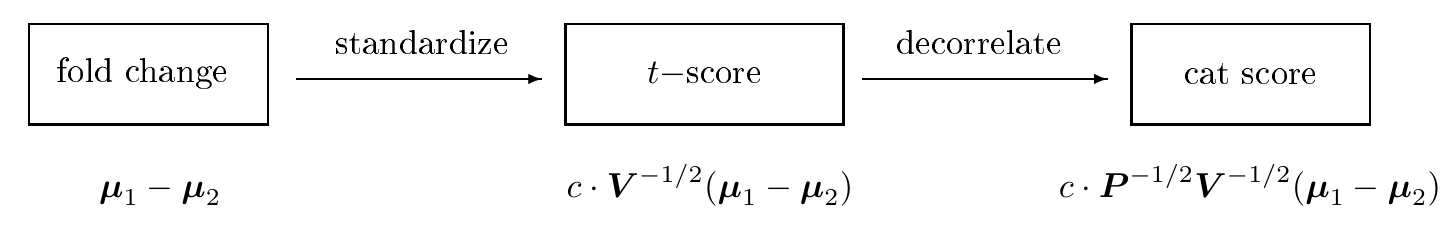}}
\caption{Relationship between fold change, $t$-score, and the cat score.
The constant $c$ equals $(\frac{1}{n_1}+\frac{1}{n_2})^{-1/2}$. }
\label{fig:catscore}
\end{center}
\end{figure*}

Using the above we define the vector $\btau^{adj}$ of
``correlation-adjusted $t$-scores'' (``cat score'') 
to be proportional to the feature weight vector $\bomega$ (\eqcite{eq:featureweights}):
\begin{equation}
\begin{split}
\btau^{adj} &\equiv (\frac{1}{n_1}+\frac{1}{n_2})^{-1/2} \, \bomega  \\
               &= \bP^{-1/2} \times \{(\frac{1}{n_1}+\frac{1}{n_2}) \bV \}^{-1/2} ( \bmu_1 -  \bmu_2 ) \\
               &= \bP^{-1/2} \btau .
\end{split}
\label{eq:catscore}
\end{equation}
Note the scale factor $(\frac{1}{n_1}+\frac{1}{n_2})^{-1/2}$ ensures that
 the empirical version of the 
cat score matches the scale of the empirical $t$-score.
The vector $\btau$ contains the gene-wise $t$-scores, and
$n_k$ is the number of observations in group $k$.

The cat score is a natural and intuitive extension of  both the
fold change and $t$-score, as illustrated in \figcite{fig:catscore}.
While the $t$-score is the standardized mean difference  $\bmu_1 -  \bmu_2$,
 the cat score is the standardized as well as \emph{decorrelated} 
mean difference.  The factor $\bP^{-1/2}$ responsible for the decorrelation
is well-known from the Mahalanobis transform that is frequently applied to 
prewhiten multivariate data. 
Also note that the inverse correlation matrix is closely 
related to  partial correlations. 

\subsection{Estimation of feature weights and computation of the cat score from data}

Substituting empirical estimates for means, variances, and correlations
into \eqscite{eq:featureweights}{eq:catscore}  provides a simple recipe for
estimating the feature
weights and computing the cat score from data.  However, this is only a valid
approach if sample size is large compared to the dimension.

For small-sample yet high-dimensional settings  we suggest 
to 
employ James-Stein-type shrinkage estimators of correlation \citep{SS05c} and of
variances \citep{OS07a}.
Plugging these two James-Stein-type estimators into \eqcite{eq:catscore} 
yields a shrinkage version of the cat score
\begin{equation}
\bt^{\text{adj}}_{\text{shrink}} =  (\bR^\text{shrink})^{-1/2} \; \bt^\text{shrink}.
\label{eq:shrinkcat}
\end{equation}

A major obstacle in the application of \eqcite{eq:shrinkcat} is the
problem of efficiently computing $(\bR^\text{shrink})^{-1/2}$. 
Direct calculation of the matrix
square root, e.g., by eigenvalue
decomposition, is extremely tedious for large dimensions $p$.
Instead, we present here a simple time-saving identity for computing 
the $\alpha$-th power of $\bR^\text{shrink}$ 
(though here we only need the case $\alpha = -1/2$).

The shrinkage correlation estimator of \citet{SS05c} is given by
$\bR^\text{shrink} = \gamma \bI_p + (1-\gamma) \bR$, 
where $\bR$ is the empirical correlation matrix and $\gamma$
the shrinkage intensity.  We define 
$\bZ = \bR^\text{shrink} / \gamma = \bI_p + \frac{1-\gamma}{\gamma} \bR  = \bI_p + \bU \bM \bU^T$, 
where $\bM$ is a symmetric
positive definite matrix of size $m$ times $m$ and $\bU$ an orthonormal basis.
Note that $m$ is the rank of $\bR$.  Subsequently, 
to calculate the $\alpha$-th power of $\bZ$
we use the identity\footnote{The validity of the identity can be verified
by noting that the eigenvalues of $(\bI_p + \bU \bM \bU^T)^\alpha$ and of the righthand side
of \eqcite{eq:matid} are identical (which implies similarity between the two matrices) and that no further rotation 
is needed for identity.
}
\begin{equation}
\bZ^\alpha = \bI_p - \bU (\bI_m -(\bI_m + \bM)^\alpha) \bU^T 
\label{eq:matid}
\end{equation}
that requires only the computation of the $\alpha$-th power of the  matrix
$\bI_m + \bM$.  This trick enables substantial computational savings when the number
of samples (and hence $m$) is much smaller than $p$.

We note that identity \eqcite{eq:matid} is related but not identical to the 
well-known Woodbury matrix identity for the inversion of a matrix.
For $\alpha=-1$ our identity reduces to
$$
\bZ^{-1} = \bI_p - \bU (\bI_m -(\bI_m + \bM)^{-1}) \bU^T,
$$
whereas the Woodbury matrix identity equals
$$
\bZ^{-1} = \bI_p - \bU (\bI_m + \bM^{-1})^{-1} \bU^T .
$$

Finally, additional information about the structure of the
correlation matrix $\bP$ (or its inverse) may also be 
taken into account when estimating the
cat score.  This is done simply by replacing
the unrestricted shrinkage estimator 
 by a more structured estimator
\citep[e.g.,][]{TP07,LL08,GLTF08}.

\subsection{Selection of single genes}

The cat score offers a simple approach to feature selection, both of 
individual genes and of sets of genes (see below).

By construction, the cat score is a decorrelated $t$-score. 
As such it measures  the individual
contribution of each single feature to separate the two groups, 
after removing the effect of all other genes.
Therefore, to select individual genes according to their relative
effect on group separation one simply ranks them according to the 
magnitude of the respective $\tau^{\text{adj}}_i$.

For determining $p$-values and FDR values we fit
a two-component mixture model to the observed cat scores \citep{Efr08a}.
Asymptotically, for large dimension  the 
null distribution is approximately normal as a consequence of
central limit theorems for dependent random variables \citep[e.g.,][]{HR48,RoWo00}
-- recall that the  cat score is a  weighted sum over $p$  
dependent $t$-statistics.  This is validated empirically in
section 3.5 discussing the analysis of a metabolomic data set.

For practical analysis we suggest employing the ``fdrtool'' 
algorithm \citep{Str08b,Str08c}.  
A comparison of methods for assigning significance to cat scores
is given in \citet{AhS09}.

\subsection{Selection of gene sets}

For evaluating the total effect  of a set of features
on group separation we exploit the close connection
of cat scores with the Hotelling's $T^2$ statistic,
 a standard criterion in gene set analysis 
\citep{LL+05,KPP06}.

Specifically,  $T^2 = (\bt^{\text{adj}})^T \bt^{\text{adj}} = \bt^T  \bR^{-1}   \bt$,
where $\bR$ is the empirical correlation matrix, $\bt^{\text{adj}}$
the empirical cat score vector, and $\bt$ the vector containing
the gene-wise Student $t$-statistic.
In other words, the $T^2$ statistic is identical to the sum of the squared 
individual empirical cat scores for the genes in the set.  Note that any normalization
with regard to the size of the set is implicit in the factor $\bR^{-1}$. For example,
if there is strong correlation among the genes in a set then $T^2$ is approximately 
the average of the underlying squared $t$-scores.

With this in mind, we define the grouped cat score for gene $i$
belonging to a given gene set as the signed square root of the
sum over the squared cat scores of all genes in the given gene set,
$$
\tau^{\text{adj,grouped}}_i = \text{sign}( \tau^{\text{adj}}_i  )   
 \sqrt{ \sum_{g \in \text{gene set}} (\tau^{\text{adj}}_g )^2 } \, .
$$

There are two main cases when it is important to consider
sets of genes rather than individual genes:
\begin{itemize}
\item First, in a gene set
enrichment analysis where \emph{prespecified} pathways or functional units
rather than individual genes are being investigated 
 (cf. \citet{AS09}). 
\item Second, if genes
are highly correlated and thus provide the same 
information on group separation.  To accommodate for this
collinearity  we suggest
constructing a suitable correlation neighborhood 
around each gene, e.g., by the rule $|r| \geq 0.85$.
Typically, the resulting sets are rather
small and for most genes are comprised only of the
gene itself -- see \citet{TW06} and also \citet{LHRG09} for similar procedures.
Note that the suggested threshold of $0.85$ -- which we use throughout this paper --
is rather conservative. It defines 
a priori which pair of genes are assumed to be collinear.
\end{itemize}
We note that using the grouped cat score 
provides to a simple procedure for high-dimensional feature selection 
where
whole sets of variables are simultaneously included or excluded,
in contrast to the classical view of feature selection where only 
one of those features is retained (see also \citet{BR08}
for references to related approaches).

\section{Results}

In order to study the performance of the cat score for 
feature selection and gene ranking, we conducted an extensive
study.   Specifically, we investigated six different correlation
scenarios, three synthetic models and three empirical correlation
matrices estimated from three different gene expression data sets,
and compared the results with  a diverse number of 
regularized $t$-scores. Furthermore, we analyzed a metabolomic
data set investigating prostate cancer.

\subsection{Correlation scenarios}

\begin{figure*}[!t]
\begin{center}
\centerline{\includegraphics[width=\textwidth]{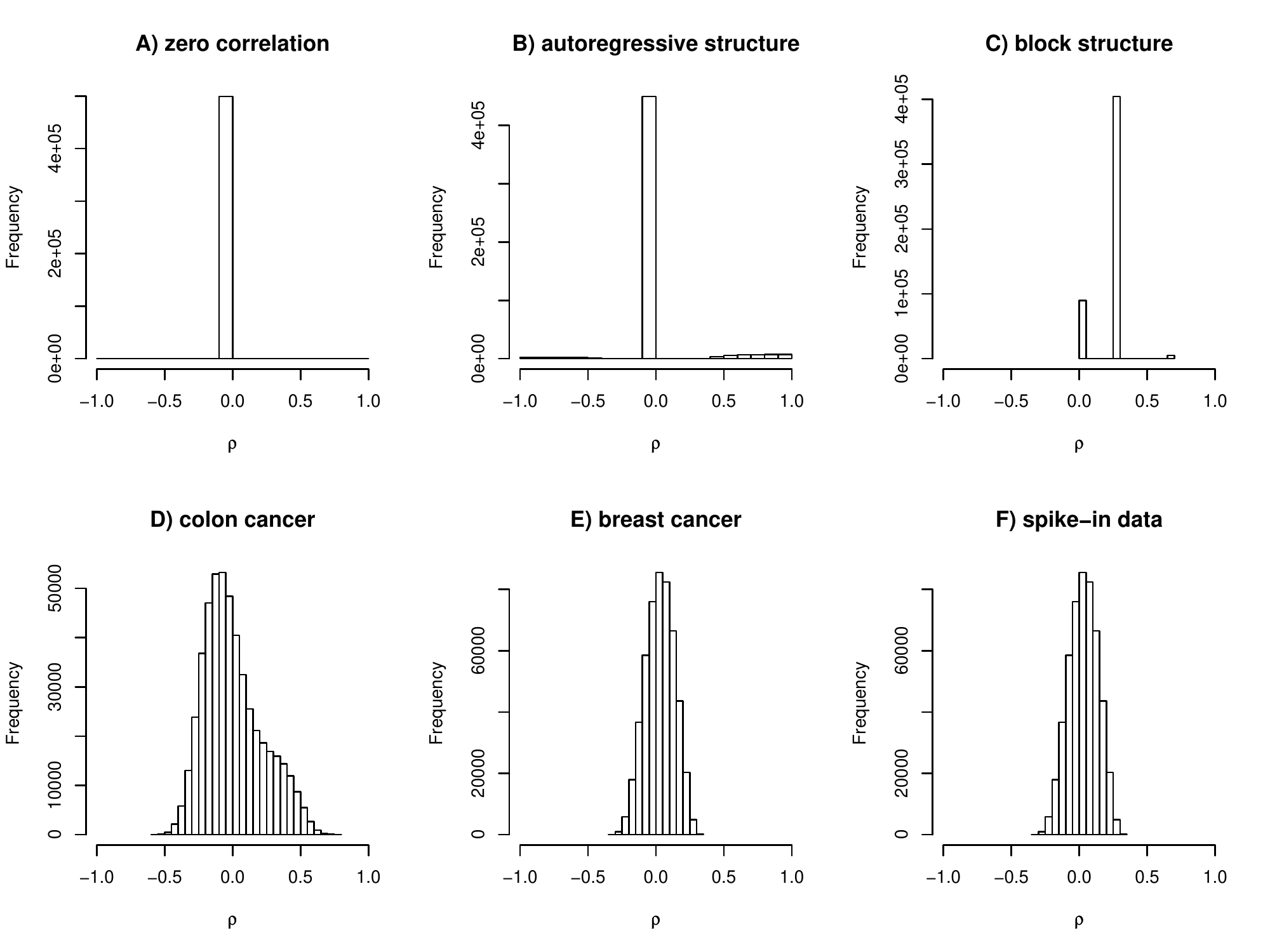}}
\caption{The six correlation scenarios investigated in our study.
All correlation matrices have size $1000 \times 1000$ and thus contain
499500 correlation values.  
\emph{Top row:} Histograms of the correlations of three synthetic correlation 
patterns (A--C).  \emph{Bottom row:} Histograms of the three empirical correlation
structures (D--F).  For further details see main text.}
\label{fig:histcorr}
\end{center}
\end{figure*}

For the correlation structure, we considered a variety of scenarios.
Specifically, we employed six different correlation patterns (cf. \figcite{fig:histcorr}):
\begin{itemize}
\item[A:]  First, as a negative control we assumed a diagonal correlation matrix
       $\bP = \bI$ of size $1000 \times 1000$.
\item[B:]  Next, we employed an
       autoregressive block-diagonal
      correlation matrix  \citep{GHT07}.  We used 10 blocks of size $100 \times 100$
      genes.   Within each block, the correlation between two genes
     $i, j, = 1, \dots, 100$ equals $\rho(i,j) = \rho^{\text{abs}(i-j)}$.
      We set $\rho=0.99$ with alternating sign in each block.
      This correlation matrix is sparse with most entries being
      very small, nevertheless it also contains some highly correlated
      genes.
\item[C:] Third, we employed a correlation block structure where the first
          100 genes have pairwise correlation of 0.7 and the remaining 900 genes
          have pairwise correlation of 0.3.  Between the two groups there is no
          correlation.  The block with the larger correlation corresponds to the
          differentially expressed genes.
\item[D:] In addition to the three artificial correlation structures, we also employed
          shrinkage estimators of correlations matrices from three expression data sets, 
          using a sample of 1000 genes.  Structure D is obtained from 
          gene expression data of colon cancer \citep{AB+99}.
\item[E:] As D, but for breast cancer  \citep{HDC+01}.
\item[F:] As D, but from a spike-in experiment \citep{CB+05}.
\end{itemize}

\subsection{Test statistics}

In our comparison we included the following gene ranking statistics:
fold change, empirical $t$ statistic,
``SAM'' \citep{TTC01}, ``moderated $t$'' \citet{Smy04}, and
``shrinkage $t$'' \citet{OS07a}.  As in \citet{OS07a} 
the latter three regularized $t$-scores gave nearly identical estimates
and always outperformed Student $t$,
so we report here only the results for ``shrinkage $t$''.
As baseline reference we also included
random ordering in the analysis.

For the cat score we investigated two variants:  the shrinkage cat score
(\eqcite{eq:shrinkcat})
and an oracle version, which uses the true underlying
correlation matrix rather than estimating the correlation structure.
For the two structures with high correlations (B and C) we employed
the grouped cat score using a correlation neighborhood threshold of 0.85.  

In addition, we included in our study two recently proposed gene
ranking procedures that, like the cat score, also aim at incorporating
information about gene-gene correlations in gene ranking:
the ``correlation-shared $t$-score''
introduced by \citet{TW06} and the ``correlation-predicted $t$-score''
suggested by \citet{Lai08}.   Correlation-shared $t$ 
averages over gene-specific Student $t$-scores in
a data-dependent correlation neighborhood. The approach by
\citet{Lai08} employs a local smoothing approach
to ``predict'' the $t$-score of a particular gene from
$t$-scores of other genes highly correlated with it.  Here,
we use the \citet{Lai08} approach with the smoothing parameter
set to its default value $f=0.2$.
Note that the cat score, the correlation-shared $t$-score and 
the correlation-predicted $t$-score all are based on linear combinations
of $t$-scores, albeit with different weights.

\subsection{Data generation}

 In our data generation procedure 
we followed closely the setup in \citet{Smy04} and \citet{OS07a}, with 
the additional specification of a correlation structure among genes.
In detail, the simulations  were conducted as follows:
\begin{itemize}
\item The number of genes was fixed at $p=1000$. The first 100 genes were
      designated to be differentially expressed.
\item The variances across genes were drawn from a 
      scale-inverse-chi-square distribution 
      $\text{Scale-inv-}\chi^2(d_0, s_0^2)$.  We used $s^2_0=4$
      and $d_0=4$, which corresponds to the ``balanced'' variance case
      in  \citet{Smy04}. Thus, the variances vary
      moderately from gene to gene.
\item The difference of means for the differentially expressed genes (1--100)
      were drawn from a  normal distribution with mean zero and the gene-specific
      variance.  For the non-differentially expressed genes (101--1000) the difference
      was set to zero.
\item The data were generated by drawing from
      group-specific multivariate normal distributions with the given 
      variances and means.  The correlation matrix assumed one of
       the above structures A--F.   
\item  We also varied the sample sizes $n_1$ and $n_2$ in each group,
       from very small $n_1=n_2=3$ to fairly large $n_1=n_2=50$.  Here,
      we report  results for  $n_1=n_2=8$.    
\end{itemize}

\subsection{Comparison of gene rankings}

\begin{figure*}[!t]
\begin{center} 

 \resizebox{\textwidth}{!}{
     \includegraphics{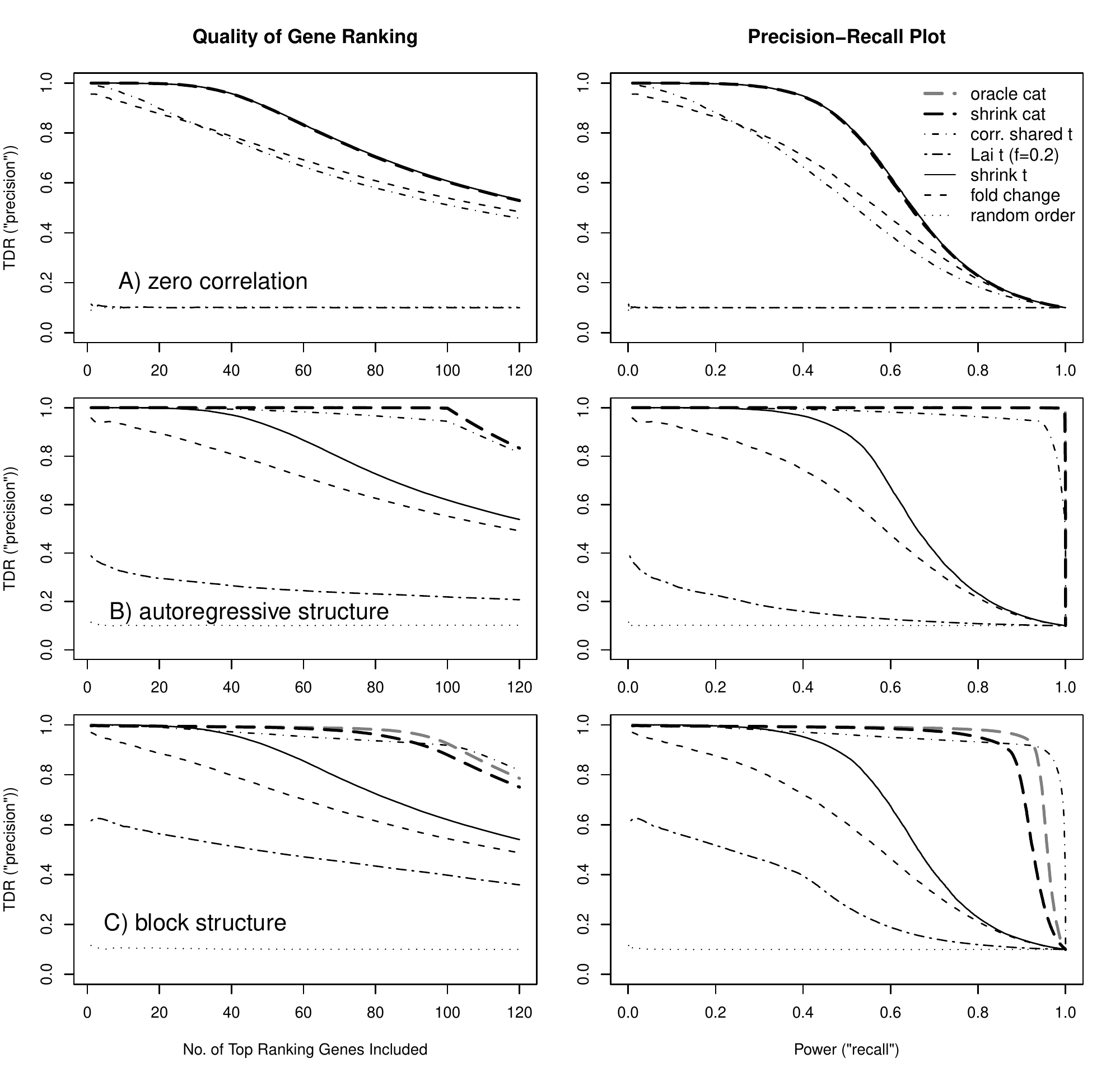}
  }   
  \caption{True discovery rates (\emph{left column}) and precision-recall curves 
(\emph{right column}) for the three synthetic correlation structures A--C.
Note that for B and C the grouped cat score was employed, using a correlation neighborhood $|r| \geq 0.85$.}

\label{fig:simulations1}

\end{center}
\end{figure*}

\begin{figure*}[!t]
\begin{center} 

 \resizebox{\textwidth}{!}{
     \includegraphics{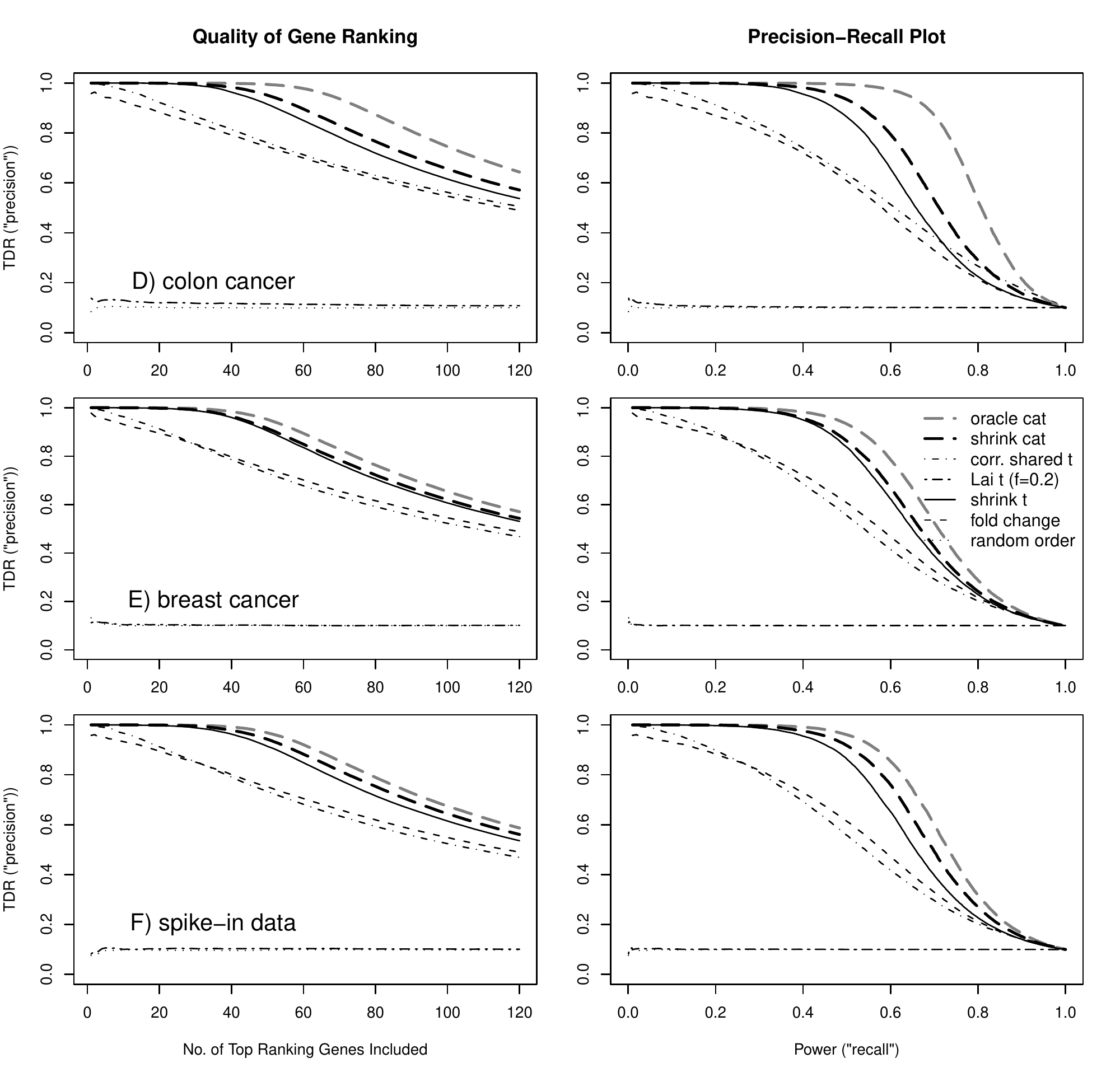}
  }   
  \caption{True discovery rates (\emph{left column}) and precision-recall curves 
(\emph{right column}) for the three empirical correlation scenarios D--F.}
\label{fig:simulations2}

\end{center}
\end{figure*}

For each correlation scenario A--F we generated 500 data sets and computed 
corresponding gene rankings using the various $t$-scores and cat scores
discussed above.  We then counted false positives ($FP$),
true positives ($TP$), false negatives ($FN$), and true negatives 
($TN$) for all possible cut-offs in the gene list (1-1000).
From this data we estimated  the true discovery rates 
 (= positive predictive value, ppv) $E( \frac{TP}{TP+FP})$
and the power (= sensitivity) $E(\frac{TP}{TP+FN})$.

A graphical summary of the results are presented in \figcite{fig:simulations1}
and \figcite{fig:simulations2}.
The first column shows the true discovery rates as a function of the
number of included top-ranking genes, whereas the second column gives
the plots of true discovery rate versus power.  The latter graphs, known in 
the machine learning community as ``precision-recall'' plots, highlight
methods that simultaneously have large power and large true discovery rates.

The first row in \figcite{fig:simulations1} shows the control case when there is no correlation present.    
As expected, the cat score performs identical to the
shrinkage $t$ approach.  
A similar performance is given by
the correlation-shared $t$ and the fold change 
statistic, slightly worse than shrinkage $t$- and cat score.  
The ordering provided by the correlation-predicted $t$-score is random, which
is not surprising as prediction fails when there is no correlation.

For the autoregressive and the block structure (scenarios B and C in
\figcite{fig:simulations1}) substantial gains are achieved
over the shrinkage $t$-score, both by the cat score and the
correlation-shared $t$-score by \citet{TW06}. In particular in case B
these two methods show near-perfect recovery of the gene ranking.
The shrinkage $t$ approach and fold change remain the second and third
best feature ranking approach, with the correlation-predicted $t$-score
of \citet{Lai08} trailing the comparison.  

For the empirically estimated correlation structures the picture changes
slightly (cf. \figcite{fig:simulations2}).  All these scenarios have in common
that there is common background correlation but no very strong individual
pairwise correlations exist  (cf. \figcite{fig:histcorr}, bottom row).
In this setting the shrinkage cat score also improves over the shrinkage $t$-score.
The oracle cat score shows that further benefits are possible if the correlation
structure was known, or if a better estimator was used.  For the empirical
 scenarios the correlation-shared $t$-score performs similar
as the fold change, and the correlation-predicted $t$-score again delivers
random orderings. 

In summary, in all the six quite different correlation scenarios
the (grouped) cat score offers in part substantial performance improvements 
over standard regularized $t$-scores, which were represented here by shrinkage
$t$-score.  The correlation-shared $t$-score also performs exceptionally well 
if there are a few highly correlated genes, but otherwise falls back to 
the efficiency of using fold-change approach.  The correlation-predicted approach
did in general not provide any reasonable orderings. 
It seems to us that this is due to the fact that it is the only test statistic
that discards the actual value of the $t$-score of a gene, 
and instead relies exclusively on closely correlated genes -- which may not exist.

\subsection{Ranking of metabolomic markers of prostate cancer}

\begin{figure*}[!t]
\begin{center} 

 \resizebox{8cm}{!}{
     \includegraphics{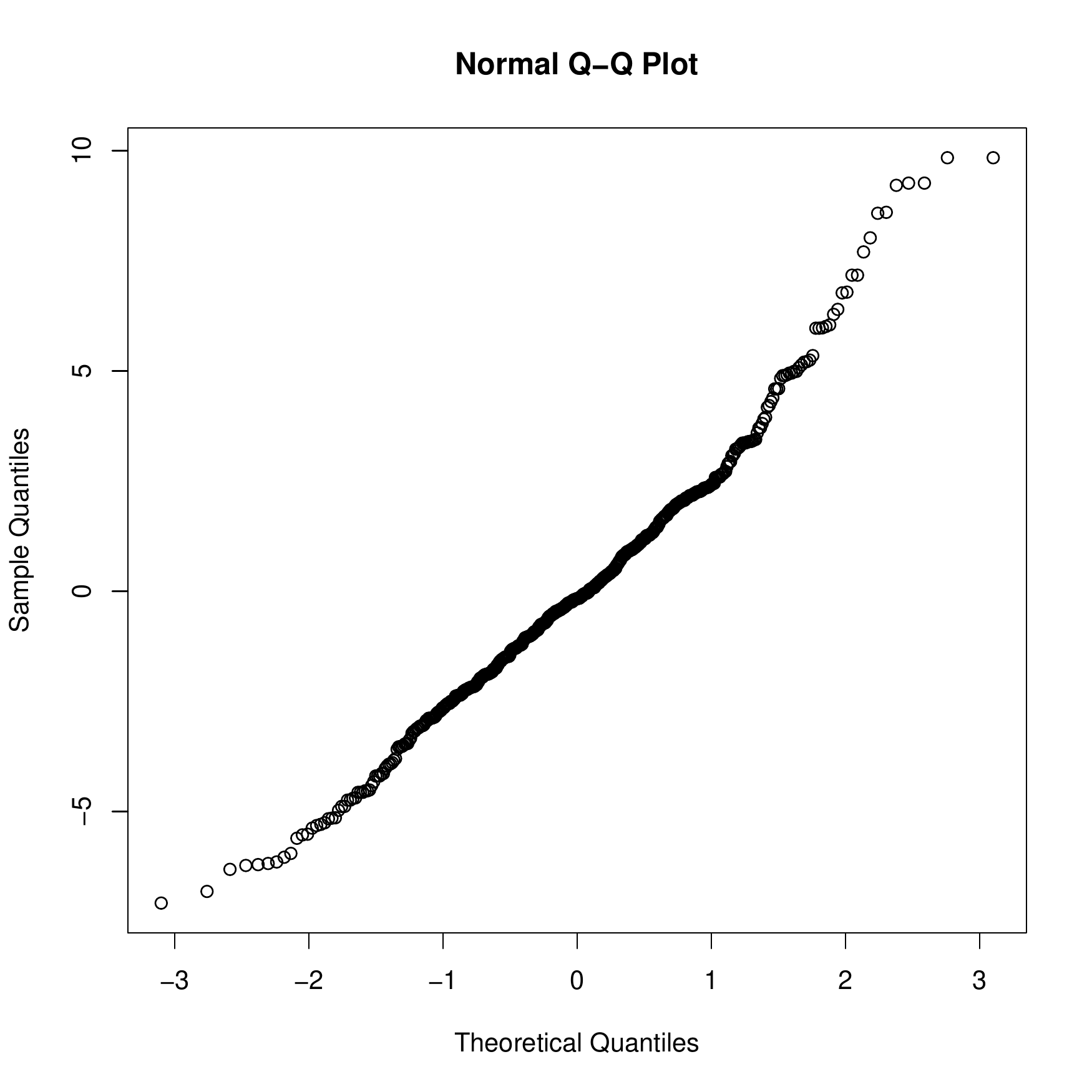}
  }   
  \caption{Plot of normal versus empirical quantiles for the grouped cat
     scores computed from the metabolomic prostate data.  The linearity
     in the central part indicates a normal null model.
}
\label{fig:qqplot}

\end{center}
\end{figure*}

To illustrate the effect of correlation on gene ranking we analyzed a subset 
of data from a recent metabolomic study concerning prostate cancer \citep{SP+09}.
The original study investigated three groups of tissues, benign, localized
cancer and metastatic prostate cancer.  Here, we focused on 
 the two types of cancer tissue.
Specifically, we compared 12 samples
of clinically localized prostate cancers versus 14 samples of
metastatic prostate cancers.  For each sample the concentrations of 
518 metabolites were measured.  We use here the preprocessed data as
kindly provided by Dr. Sreekumar and Dr. Chinnaiyan.

For each of the 518 metabolites we computed a shrinkage $t$-score and
a shrinkage cat score. For the latter we applied grouping of features
with a correlation threshold of $|r| \geq 0.85$.   The Q-Q-plot of the cat
scores versus a normal distribution is shown in \figcite{fig:qqplot}. 
By inspection of this diagnostic plot we see that the 
null model of the grouped cat scores, represented by
the linear middle part, is approximately
a normal distribution.
The deviations from normality at the tails correspond to the
alternative distribution containing the high-ranked metabolites of interest.

\begin{table}[!b]
\caption{The top 10 ranking metabolites according to the shrinkage $t$ and 
the grouped cat scores, respectively. Note that nicotinamide and X-5207, as
well as guanosine and X-3390, are strongly
correlated.}
\centering
\begin{tabular}{lrr}
\toprule
Rank         & shrinkage $t$ & grouped cat score\\
\midrule
1 & Ciliatine    & Nicotinamide \\
2 & Inosine      & X-5207 \\
3 & Putrescine   & Guanosine \\
4 & X-3390       & X-3390 \\
5 & Palmitate    & Ciliatine \\
6 & Glycerol     & Putrescine \\
7 & Ribose       & Inosine \\
8 & X-3102       & Citrate \\
9 & Myristate    & Uridine \\
10 & X-4620      & X-2867 \\
\bottomrule
\end{tabular}
\label{tab:metabolomic.data}\\

\end{table}

The ten top ranking metabolic features that differentiate between
localized and metastatic cancer according to $t$-scores and cat scores,
respectively,
are listed in \tabcite{tab:metabolomic.data}.
Overall, the two rankings differ quite notably, as expected in
the presence of correlation.  In particular, at the
top of the list there are differences
due to very strong correlation between the substrate X-5207 and 
Nicotinamide ($r =  0.9444$) and likewise between  Guanosine and X-3390
($r= 0.9389$).  Unlike with $t$-scores, in a grouped cat score analysis the
features in these two pairs are treated as a unit.  Jointly, the correlated
markers outperform other individual markers with respect to distinguishing
between
the two phenotypic groups. 

Regarding the interpretation of observed enrichment of
nicotin\-amide and guanosine, we caution that 
without further additional information it is not possible to
decide whether this is due to intake
of medication or rather due to the different progression of cancer.

For a series of other data examples further illustrating
the analysis of cat scores
and estimation of corresponding predictive errors
we refer to \citet{AhS09}.

\section{Discussion}

\subsection{Harmonizing gene ranking and feature selection}

The correlation-adjusted $t$-score is the result of our attempt to 
harmonize gene ranking with LDA feature selection.   While it is well
known that in the absence of correlation the $t$-score
provides optimal rankings \citep{FF08}, the situation is less clear
in the LDA case where genes are allowed to be correlated.
Here we show that the cat score provides a natural weight for
feature selection in LDA analysis and that it can be sucessfully employed to 
rank genes and gene groups.

In order to apply cat scores in the analysis of high-dimensional data 
we develop in this paper a corresponding shrinkage procedure.  For moderately
high dimensions and sufficient sample size we demonstrate that incorporating
correlation information into the gene ranking can lead to substantial
improvement in power.   However, this is only feasible
if either the sample size is large or the signal is strong enough
to estimate correlations \citep{HMN05}.   For microarray data with 
very small sample size (in the order of $n_1=n_2=3$) it is impossible
to estimate a large-scale correlation matrix, and unsurprisingly
for that case we did not see any benefits.  However, as our study shows (\figcite{fig:simulations1}
and \figcite{fig:simulations2}) using the cat score can lead to substantial
gains already for relatively moderate sample sizes ($n_1=n_2=8$).

\subsection{Recommendations}

In high-dimensional genomic experiments with very small sample size,
when  nothing is known a priori about the correlation structure, 
we recommend employing the standard regularized $t$-scores.

However, for moderate ratios of $p/n$, say smaller than 100, 
it is often possible to
obtain reliable estimates of the correlation among markers.
Thus, in this setting we propose ranking of biomarkers by the 
correlation-adjusted $t$-score, computed by the shrinkage procedure
outlined above.  In addition, if
inspection of the correlation histogram shows existence of
highly correlated features, then joint evaluation of those features 
by computing the grouped cat score is advised.  Using
more constrained correlation estimators may further improve
the efficiency.

Finally, as pointed out by a referee, gene ranking by cat 
scores may  be combined with fold change-based thresholding,
in order to filter out statistically significant yet biologically
irrelevant features  \citep[e.g.][]{MS09}.

In short, we propose to view gene ranking as a generically multivariate problem.
In this perspective it seems stringent not only to  standardize the mean differences
(i.e.\ using the corresponding $t$-scores) but also to additionally 
 decorrelate them, which results in the cat score proposed here.

\section*{Acknowledgments}
We are grateful to the anonymous referees for their very valuable
comments. We thank our colleagues at IMISE for discussion
and  Anne-Laure Boulesteix, Florian
Leitenstorfer and Abdul Nachtigaller for additional suggestions.

\appendix

\section*{Appendix: Computer implementation}

The ``shrinkage cat'' estimator
(\eqcite{eq:shrinkcat}) is implemented in the R  package ``st'', which 
is freely available under the terms of the GNU General Public License
(version 3 or later) from CRAN
(\url{http://cran.r-project.org/})
and from URL \url{http://strimmerlab.org/software/st/}.